\documentclass[10pt,aps,prd,twocolumn,showpacs,amsmath,amssymb,nofootinbib,eqsecnum,preprintnumbers,superscriptaddress]{revtex4-2}
\usepackage{amsmath,amssymb}
\usepackage{enumitem}  
\usepackage[T1]{fontenc}
\usepackage[usenames, dvipsnames]{color} 
\usepackage{graphicx} 
\usepackage{comment}
\usepackage[normalem]{ulem}
\usepackage[utf8]{inputenc}
\UseRawInputEncoding 
\usepackage{url}
 \usepackage{xcolor}
 \usepackage{multirow}
 \usepackage{stix} 
\usepackage{bm}
\usepackage{amsthm}
\newcommand{\mathds}[1]{\text{\usefont{U}{dsrom}{m}{n}#1}}

\usepackage{hyperref}

\newcommand{\be}{\begin{equation}}
\newcommand{\ee}{\end{equation}}
\newcommand{\ba}{\begin{eqnarray}}
\newcommand{\ea}{\end{eqnarray}}

\newcommand{\beq}{\begin{equation}}
\newcommand{\eeq}{\end{equation}}
\newcommand{\beqa}{\begin{eqnarray}}
\newcommand{\eeqa}{\end{eqnarray}}

\newcommand{\im}{\mathrm{i}}

\newcommand{\tho}{\text{\normalfont\th}}      
\newcommand{\edt}{\eth}                       
\newcommand{\Pb}{\bar\Phi}
\newcommand{\anp}{\alpha_{\rm NP}}
\newcommand{\enp}{\varepsilon_{\rm NP}}
\newcommand{\anpb}{\bar{\alpha}_{\rm NP}}
\newcommand{\enpb}{\bar{\varepsilon}_{\rm NP}}
\newcommand{\fe}{\mathfrak e}
\newcommand{\frg}{\mathfrak g}
\newcommand{\sD}{\mathscr D}
\newcommand{\sA}{\mathsf A}
\newcommand{\sF}{\mathsf F}
\newcommand{\sJ}{\mathsf J}
\newcommand{\sX}{\mathsf X}
\newcommand{\sC}{\mathsf C}
\newcommand{\sK}{\mathbb K}
\newcommand{\sU}{\mathbb U}
\newcommand{\sz}{\mathsf z}
\newcommand{\pa}{\partial}
\newcommand{\Vq}{\mathcal V_q}
\newcommand{\Vp}{\mathcal V_p}

\begin{document}

\title{Separability of coupled gravitational perturbations in Pleba\'nski--Demia\'nski spacetimes}%

\author{Finnian Gray}

\email{finnian.gray@mq.edu.au}
\affiliation{School of Mathematical and Physical Sciences, Macquarie University, NSW 2109, Australia}

\author{Hryhorii Ovcharenko}

\email{hryhorii.ovcharenko@matfyz.cuni.cz}

\affiliation{Institute of Theoretical Physics, Faculty of Mathematics and Physics,
Charles University, Prague, V Hole{\v s}ovi{\v c}k{\' a}ch 2, 180 00 Prague 8, Czech Republic}

\date{23rd November, 2026}

\begin{abstract}
Recently, a major breakthrough in the separability of gravitational perturbations of the Kerr--Newman rotating charged black hole was achieved by Hintz [\href{https://arxiv.org/pdf/2609.33661}{\sf arxiv:2609.33661}], and extended to non-zero cosmological constant Kerr--Newman-(A)dS black holes by Yoshida [\href{https://arxiv.org/pdf/2610.20500}{\sf arxiv:2610.20500}] and Zhao [\href{https://arxiv.org/pdf/2610.22274}{\sf arxiv:2610.22274}]. 
This method removes the need for the freezing approximation, wherein coupling to the electromagnetic degrees of freedom are ignored, at the price of a coupled set of equations in the gravitational radiative degrees of freedom.
The key insight is to transform the coupled system of equations into a chiral form which is amenable to a modified matrix $R$-separation of variables. 
In this work we extend the construction to the most general type D electro-vacuum solution to the Einstein--Maxwell equations, with aligned matter and cosmological constant.
The construction makes use of the conformal properties of the equations and we find a symmetry operator underlying the decoupling of the chiral system.

\end{abstract}

 \maketitle

\section{Introduction}
Black hole perturbation theory is one of the key tools for studying the gravitational radiation ~\cite{Sasaki:2003xr,Pound:2021qin}. 
Perturbation theory is most useful when there is a possibility performing a separation of variables for the gravitational radiation which allows for the analysis of linear mode stability e.g.~\cite{Andersson:2019dwi}.
This problem is however not straightforward, it took ten years from the discovery of the Kerr metric describing asymptotically flat rotating black holes~\cite{Kerr:1963ud} until its gravitational perturbations were able to be effectively studied using Teukolsky's major breakthrough~\cite{Teukolsky1973,Teukolsky:1972my,Teukolsky:1974yv}. 
Teukolsky showed that the curvature perturbations, in contrast to metric ones, satisfy a wave equation which admits a separation of variables.

Underpinning these results are the so-called hidden symmetries of black hole spacetimes in general relativity. 
In particular, these are represented by Killing tensors and Killing--Yano (KY) forms~\cite{Yano}.
First used to study the separability of geodesics~\cite{Carter:1968rr}, they govern the fundamental properties of the general class of Kerr--NUT--AdS spacetimes in four \cite{carter1968new, Carter:1968ks} and higher \cite{Chen:2006xh} dimensions. 
Hidden symmetries are the origin of the separability of Hamilton--Jacobi, Klein--Gordon, Dirac, and (massive) vector equations in these spacetimes, enforce the type algebraic type D of the Weyl tensor of the solution, and lead uniquely to the (off-shell) Kerr--NUT--AdS geometry\cite{Krtous:2008tb}.
See the review \cite{Frolov:2017kze} and references therein for more details. 
Tied to the underlying structure of the Teukolsky equation is the Newman--Penrose formalism~\cite{Newman:1961qr} and its extension by Geroch, Held, and Penrose~\cite{Geroch:1973am}.
This is intimately tied to the 2-spinor formalism~\cite{PenroseRindler1,PenroseRindler2} where the KY tensor is represented as a Killing Spinor~\cite{Walker:1970un}.
This construction has also been used to study symmetries operator, test fields, and separation of variables, see for example~\cite{Andersson:2014lca,Andersson:2015xla,Araneda:2016iwr,Aksteiner:2021nec}.

In four dimensions, many of these properties carry over, in a slightly weaker conformal sense, to Pleba\'nski--Demia\'nski (PD) spacetimes, which possess a conformal Killing--Yano  (CKY) tensor~\cite{Kubiznak:2007kh} or equivalently a conformal Killing spinor.
These describe the most general type D spacetime with aligned electromagnetic fields~\cite{Plebanski:1976gy}. 
In addition to being a rotating charged black hole, with a cosmological constant and NUT parameter, the PD spacetime contains an additional parameter corresponding to a uniform acceleration, due to tension from an unbalanced cosmic string.

However, the extension of separability of gravitational perturbations to charged rotating black holes turns out to be problematic. 
The perturbations of the Maxwell scalars mix with the curvature perturbations and the system does not yield a decoupled wave equation~\cite{Chandrasekhar1998}.
The problem remains in PD spacetimes.
To make progress, a ``freezing'' approximation is typically used which effectively ignores this mixing of electromagnetic and gravitational degrees of freedom. 
Dudley and Finley ~\cite{dudley1977} were able to use this approximation to construct a separable wave equation for the Teukolsky variables, whose origin lies in the conformal properties of the Teukolsky equation~\cite{Araneda:2018ezs}, (see also \cite{Gray:2025lwy}).
Although this approximation is well used~\cite{saha2025}, it remains unsatisfactory, especially for proving mode stability of gravitational perturbations, as the terms dropped are of the same order as the terms kept.

Quite remarkably this obstacle was overcome only very recently (another 47 years after the work of Dudley and Finley) by Hintz~\cite{Hintz:2026jiw} for Kerr--Newman black holes. 
Quickly this was extended to non-zero cosmological constant by Yoshida~\cite{Yoshida:2026yly} (and Zhao independently later \cite{Zhao:2026gen}).
The technique and resulting separated equations have found immediate applications, see e.g. \cite{Destounis:2026ljd,Weller:2026bfn,Chakraborty:2026kyf,Sarkar:2026xkq}
The key insight is that the coupled system can be put into a chiral form which admits a matrix $R$-separation reminiscent of works on the Dirac equation~\cite{Kalnins:1986wa,Carignano_2011}.
See also \cite{Santos:2026pbu,Dias:2026zri} for other recent separability breakthroughs.
In this work we apply the methodology to the PD metric and explicitly separate the resulting chiral system. 
In addition we identify a symmetry operator which underlies the $R$-separation techniques.

The paper is organised as follows: In Sec. \ref{Sec: PD} we present the PD line element and its hidden symmetries. 
Next, in Sec. \ref{Sec: GHP} we introduce the Geroch--Held--Penrose (GHP) extension~\cite{Geroch:1973am} of the Newman--Penrose (NP)~\cite{Newman:1961qr} formalism.
This enables efficient calculation of the gravitational perturbative system in Sec. \ref{Sec: Linear Eqs}. 
We then construct the chiral form of the equations in Sec. \eqref{Sec: chiral} and separate the equations in Sec. \ref{Sec: Separation}. 
In Sec. \ref{Sec: Sym Operator} we show they possess an inherent symmetry operator and wrap up with an outlook for future work in \ref{Sec: Outlook}.
Throughout we use mostly plus signature, lowercase Latin letters  $abcd\dots$  for spacetime indices,  (in the appendix capital Latin letters $ABC\dots$ denote spinor ones), a bold symbol $\bm k$ denotes an abstract vector, and we follow the Riemann curvature and sign conventions of \cite{Stephani:2003tm}.

\section{Pleba\'nski--Demia\'nski spacetimes}\label{Sec: PD}

The Pleba\'nski--Demia\'nski~\cite{Plebanski:1976gy} line element can be written in many different ways~\cite{Griffiths:2005qp,Podolsk2021,Podolsk2023,Ovcharenko:2024yyu,Ovcharenko:2025fxg}, however we shall employ the ``conformal-to-Carter'' form~\cite{Gray:2025lwy} also entitled in~\cite{Ovcharenko:2024yyu} as $\mathrm{PD_{\alpha}}$ metric form so that,
\begin{equation}\label{eq:  PD}
\begin{split}
ds^2=\frac{1}{\Omega^2}\Big[&-\frac Q\Sigma(d\eta-p^2d\phi)^2+\frac\Sigma Qdq^2\\
&+\frac P\Sigma(d\eta+q^2d\phi)^2+\frac\Sigma Pdp^2\Big]\,,
\end{split}
\end{equation}
with $\Sigma=q^2+p^2$ and
\begin{equation}\label{eq:  metric funcs}
\begin{aligned}
\Omega&=1-\alpha pq,\\
Q&=k+e^2-2mq+\epsilon q^2-2\alpha nq^3-(\alpha^2k+\lambda)q^4\,,\\
P&=k+2np-\epsilon p^2+2\alpha mp^3-\big(\alpha^2(k+e^2)+\lambda\big)p^4\, .
\end{aligned}
\end{equation}

The corresponding Maxwell potential is
\begin{equation}\label{eq: APD}
A=-\frac1\Sigma\big[\fe\, q\,(d\eta-p^2d\phi)+\frg\, p\,(d\eta+q^2d\phi)\big].
\end{equation}
Here $m$ is the mass, $n$ the NUT parameter, $\fe$ and $\frg$ the electric and magnetic charges with $e^2=\fe^2+\frg^2$, $\alpha$ the acceleration, $\lambda=R/12=\Lambda_c/3$ the cosmological constant (for the simplicity of calculations we introduce the  parameter $\lambda$ instead of usual cosmological constant $\Lambda_c$ simply not to have an additional factor $1/3$ in the corresponding equations), and $k$, $\epsilon$ are constants, one of which can be fixed by rescaling.
The metric~\eqref{eq:  PD} has two isometries, the Killing vectors $\pa_\eta$, $\pa_\phi$. 
The PD metric is a solution of the Einstein--Maxwell equations, which we present in NP variables in the next section.

In these coordinates the factor inside the square bracket of \eqref{eq:  PD} is the off-shell Carter spacetime~\cite{carter1968new, Carter:1968ks} which corresponds to the most general four dimensional spacetime to have a non-degenerate closed conformal Killing--Yano (CCKY) 2-form~\cite{Krtous:2008tb} and represents the Kerr--Newman--NUT-(A)dS family, for modified metric functions \eqref{eq:  metric funcs}. 
Throughout a hatted object will indicate an object in the off-shell Carter metric.
The symmetry properties of off-shell Carter metrics all descend from the CCKY tensor~\cite{Frolov:2017kze} $\hat h$, its eigenvalues are $q$ and $p$, its eigenvectors are the principal null directions, and it generates the Killing--Yano form $\hat f=\star\hat h$, the Killing tensor $\hat K_{ab}=\hat f_a^c\hat f_{bc}$ and the Killing vectors of $\hat g$ (and in higher dimensions the whole Killing tower~\cite{Frolov:2017kze}).

On the other hand, PD spacetimes only possess a conformal Killing--Yano tensor~\cite{Kubiznak:2007kh}, $h$, which although weaker provides the required structure for the separation of conformally invariant fields~\cite{Gray:2025lwy}. 
It is given by,
\begin{equation}\label{eq:  CKY}
h=\Omega^{-3}\hat h=\Omega^{-3}\,d\hat b,\qquad \hat b=\frac12\big[(p^2-q^2)\,d\eta+p^2q^2\,d\phi\big].
\end{equation}
The CKY is equivalent to a conformal Killing spinor 
and its existence alone forces the spacetime to be type D, with shear-free aligned null geodesic congruence similar to the Goldberg--Sachs theorem~\cite{Goldberg1962}, but without any field equations~\cite{PenroseRindler2} (see also appendix \ref{App: K S}).

We now introduce the necessary machinery to derive the system of equations describing radiative gravitational perturbations of the PD background spacetime.

\section{GHP formalism and curvature scalars}\label{Sec: GHP}

In this section we introduce the Geroch--Held--Penrose (GHP) extension~\cite{Geroch:1973am} of the Newman--Penrose (NP)~\cite{Newman:1961qr} formalism. 
This formalism enables efficient dealing with the conformal properties of the spacetime and hence relating the radiative perturbative system of the Carter spacetime to the PD one. 

Given a null tetrad $g({\bm k},{\bm l})=-1$, $g({\bm m},\bar{\bm m})=1$, as usual we introduce the derivative operators along the null directions $D=k^a\nabla_a$, $\Delta=l^a\nabla_a$, $\delta=m^a\nabla_a$, for which the spin coefficients are defined
\begin{equation}\label{eq: spincoef}
\begin{gathered}
\kappa=-m^aDk_a,\quad \sigma=-m^a\delta k_a,\quad \rho=-m^a\bar\delta k_a,\\ \tau=-m^a\Delta k_a,\quad
\kappa'=-\bar{m}^a\Delta l_a,\quad \sigma'=-\bar{m}^a\bar\delta l_a,\\ \rho'=-\bar{m}^a\delta l_a,\quad \tau'=-\bar{m}^aDl_a,\\
2\enp=k^aDl_a-m^aD\bar{m}_a,\quad 2\gamma=k^a\Delta l_a-m^a\Delta\bar{m}_a,\\
2\beta=k^a\delta l_a-m^a\delta\bar{m}_a,\quad 2\anp=k^a\bar\delta l_a-m^a\bar\delta\bar{m}_a .
\end{gathered}
\end{equation}

The Weyl, $C_{abcd}$, and Ricci, $R_{ab}$, curvature scalars are:
\begin{align}
    \label{eq: Weyl coeffs}
    \Psi_0&=C_{abcd}k^am^bk^cm^d, \quad \Psi_1=C_{abcd}k^al^bk^cm^d,
    \nonumber\\
    \Psi_2&=C_{abcd}k^am^b\bar{m}^cl^d\,, \quad \Psi_3=C_{abcd}k^al^b\bar{m}^cl^d\,,
    \nonumber\\
    \Psi_4&=C_{abcd}\bar{m}^al^b\bar{m}^cl^d\,,
    \\
    \label{eq: Ricci coeffs}
    \Phi_{00}&=\frac12R_{ab}k^ak^b, \quad \Phi_{01}=\frac12R_{ab}k^am^b, 
    \nonumber\\
    \Phi_{11}&=\frac14R_{ab}(k^al^b+m^a\bar{m}^b), \quad \Lambda=R/24\,.
\end{align}
In addition, the Maxwell scalars are
\begin{align}\label{eq: max scalars}
    \Phi_0&=F_{ab}k^am^b, \quad \Phi_1=\frac12F_{ab}(k^al^b+\bar{m}^am^b)\,, 
    \nonumber\\
    \Phi_2&=F_{ab}\bar m^a l^b\,.
\end{align}

In this notation the Einstein--Maxwell-$\Lambda$ equations become
\begin{equation}
    \Phi_{ij}=2\Phi_i\Pb_j\,,\Lambda=\text{const.}
\end{equation}
For the PD metric \eqref{eq:  PD}, the solution takes the form $\Lambda=\lambda/2$, where $\lambda$ is the parameter in~\eqref{eq:  metric funcs}.
So far this is the standard NP formalism. 
To reduce the number of variables the GHP formalism makes use of transformations of the NP coefficients under boosts and rotations. 
Briefly we recap this here.

Under the transformation $\bm k\to|\lambda|^2{\bm k}$, ${\bm l}\to|\lambda|^{-2}{\bm l}$ and ${\bm m}\to\lambda\bar\lambda^{-1}{\bm m}$ a quantity of weight $\{p,q\}$ scales as $\lambda^p\bar\lambda^q$, and the prime operation sends $\{p,q\}\to \{-p,-q\}$.
The weights of the spin coefficients and curvature and Maxwell scalars are 
\begin{align}
   \kappa&\sim\{3,1\}, \sigma\sim\{3,-1\}, \rho \sim \{1,1\}, \tau\sim\{1,-1\}\\
   \Psi_n&\sim\{4-2n,0\},  \Phi_{ij}\sim\{2-2i,2-2j\}, \\
   \Phi_n&\sim\{2-2n,0\}. 
\end{align}
One introduces the following operators $\tho,\eth$ which are covariant with respect to these weights. 
They take a weight $\{p,q\}$ scalar to  weights $\{p+1,q+1\}$ and  $\{p+1,q-1\}$  respectively and are given by:
\[
\begin{aligned}
\tho&=D-p\enp-q\enpb,&\tho'&=\Delta-p\gamma-q\bar\gamma,\\
\edt&=\delta-p\beta-q\anpb,&\edt'&=\bar\delta-p\anp-q\bar\beta .
\end{aligned}
\]
This plus their commutation relations, the Bianchi and Ricci identities (see App. \ref{App: Exact eqs}) neatly organise calculations.
Finally, the prime  operation sends ${\bm k}\leftrightarrow {\bm l} $, ${\bm m}\leftrightarrow \bar {\bm m}$, $\kappa\leftrightarrow\kappa'$, $\sigma\leftrightarrow\sigma'$, $\rho\leftrightarrow\rho'$, $\tau\leftrightarrow\tau'$, $\tho\leftrightarrow\tho'$, $\edt\leftrightarrow\edt'$, $\Psi_n\to\Psi_{4-n}$, $\Phi_n\to-\Phi_{2-n}$, and $\Phi_{ij}'=\Phi_{2-i\,2-j}$.
Thus, one can easily obtain negative spin weighted perturbed quantities using the prime operation.

As mentioned in section~\ref{Sec: PD}, the existence of a CKY tensor (or equivalently a conformal Killing spinor) implies the following spin-coefficients vanish $\kappa=\sigma=0$ as do the curvature scalars $\Psi_0=\Psi_1=0$, and the spacetime is type D~\cite{Walker:1970un} with shear-free null geodesics in the principal direction, regardless of the field equations (this is detailed in Appendix~\ref{App: K S}).
For the PD metric \eqref{eq:  PD} we choose the null tetrad in the form:
\begin{equation}\label{eq:  frame}
\begin{aligned}
{\bm k}&=k=\Omega^2\hat {\bm k},& \hat {\bm k}&=\frac{q^2\pa_\eta-\pa_\phi}{Q}+\pa_q,\\
{\bm l}&=\hat {\bm l},& \hat {\bm l}&=\frac{q^2\pa_\eta-Q\pa_q-\pa_\phi}{2\Sigma},\\
{\bm m}&=\Omega\hat {\bm m},& \hat {\bm m}&=\frac{1}{\sqrt2z}\Big(\frac{i(p^2\pa_\eta+\pa_\phi)}{v}+v\pa_p\Big),
\end{aligned}
\end{equation}
with $z=q+ip,\bar z=q-ip$, $u=\sqrt Q$, $v=\sqrt P$. 
In the case of $\Omega=1$ (meaning there is no acceleration $\alpha=0$), this tetrad reduces to a Kinnersley-like tetrad for Carter spacetime.

The spin coefficients of the off-shell Carter metric are
\begin{equation}\label{eq:  Carterspin}
\begin{gathered}
\hat\kappa=\hat\sigma=\hat\kappa'=\hat\sigma'=\hat\varepsilon_{\rm NP}=0,\quad \hat\rho=-\frac1{\bar z},\quad \hat\tau=\frac{iv}{\sqrt2\Sigma},\\
\hat\rho'=\frac{Q}{2\Sigma \bar z},\quad \hat\tau'=\frac{iv}{\sqrt2{\bar z}^2},\quad \hat\gamma=-\hat\rho'+\frac{Q'}{4\Sigma},\\
\hat\beta=\frac{P'}{4\sqrt2\,vz},\quad \hat\alpha_{\rm NP}=-\hat\tau'-\frac{P'}{4\sqrt2\,v\bar z},
\end{gathered}
\end{equation}
and, with $L=\ln\Omega$, the non-zero PD ones can be written: $\rho=\Omega^2(\hat\rho+\hat\tho L)$, $\tau=\Omega(\hat\tau+\hat\edt L)$, $\rho'=\hat\rho'+\hat\tho'L$, $\tau'=\Omega(\hat\tau'+\hat\edt'L)$, $\beta=\Omega\hat\beta$, $\anp=\Omega(\hat\alpha_{\rm NP}+\hat\edt'L)$, $\gamma=\hat\gamma+\hat\tho'L$, $\enp=0$. 
Explicitly these read.
\begin{equation}\label{eq:  spinexplicit}
\begin{aligned}
\rho&=-\frac{\Omega^2}{\bar z}+\Omega\Omega_{,q},&
\tau&=\frac{i\Omega v}{\sqrt2\Sigma}+\frac{v\,\Omega_{,p}}{\sqrt2z},\\
\rho'&=\frac{Q}{2\Sigma \bar z}-\frac{Q\,\Omega_{,q}}{2\Sigma\Omega},&
\tau'&=\frac{i\Omega v}{\sqrt2{\bar z}^2}+\frac{v\,\Omega_{,p}}{\sqrt2{\bar z}}.
\end{aligned}
\end{equation}
These expressions hold off shell, for arbitrary $Q(q)$, $P(p)$, $\Omega(q,p)$. 
The only non-zero Weyl scalar is
\begin{equation}\label{eq: Psi2}
\Psi_2=\Omega^2\Big[\frac{Q-P-\frac {\bar z}2(Q'-iP')}{\Sigma \bar z^2}+\frac{Q''+P''}{12\Sigma}\Big].
\end{equation}
On shell for PD, the only non-zero curvature and Maxwell scalars are
\begin{equation}\label{eq:  bgscalars}
\Phi_1=\frac{\Omega^2(\fe-i\frg)}{2\bar z^2},\quad \Psi_2=\frac{\Omega^3}{\bar z^3}\Pi,\quad \Pi=\frac{e^2(2-\Omega)}{z}-\bar\mu,
\end{equation}
where $\mu=m+in$. For $\alpha=n=0$, $k=a^2$ and $\epsilon=1-\lambda a^2$, \eqref{eq:  PD} is the Kerr--Newman--(A)dS metric in Carter coordinates.

We can now turn to discuss the radiative system.

\section{The linearised radiative system of equations}\label{Sec: Linear Eqs}

In this section we derive the linearised system of equations for the radiative gravitational degrees of freedom.
To do this we use the Ricci, Bianchi and commutator identities along with the Einstein--Maxwell equations in the GHP formalism. 
We obtain equations for the variables $\{\kappa, \sigma, \Psi_0, \Psi_1,\Phi_0\}$ and the primed objects $\{\kappa', \sigma', \Psi_4, \Psi_3,\Phi_2\}$ which vanish on the background and are therefore invariant under linearised diffeomorphisms~\cite{Stewart:1974uz}. 

To perturb the system we consider Maxwell's equations in an arbitrary spacetime
\begin{equation}\label{eq:  M}
\begin{aligned}
\tho\Phi_1-\edt'\Phi_0&=-\tau'\Phi_0+2\rho\Phi_1-\kappa\Phi_2,\\
\edt\Phi_1-\tho'\Phi_0&=-\rho'\Phi_0+2\tau\Phi_1-\sigma\Phi_2,
\end{aligned}
\end{equation}
with their primes. 
On the background they reduce to 
\begin{align}
    \tho\Phi_1&=2\rho\Phi_1, \edt\Phi_1=2\tau\Phi_1\,.
\end{align}

Inserting the Einstein--Maxwell equations $\Phi_{ij}=2\Phi_i\Pb_j$ into the Bianchi identities \eqref{eq:  Bianchi}, the derivatives of $\Pb_i$ recombine into the conjugate of the Maxwell equations which vanish on shell, and one obtains, exactly,
\begin{align}
&(\edt'-\tau')\Psi_0-(\tho-4\rho)\Psi_1-3\Psi_2\kappa\nonumber\\
&=2\Pb_0(\edt\Phi_0+2\sigma\Phi_1)-2\Pb_1(\tho\Phi_0+2\kappa\Phi_1),\label{E1}\\
&(\tho'-\rho')\Psi_0-(\edt-4\tau)\Psi_1-3\Psi_2\sigma\nonumber\\
&=2\Pb_1(\edt\Phi_0+2\sigma\Phi_1)-2\Pb_2(\tho\Phi_0+2\kappa\Phi_1).\label{E2}
\end{align}
The Ricci identity \eqref{eq:  Ricci} for $\tho\sigma-\edt\kappa$ (on an aligned type D background) is
\begin{equation}
(\tho-\rho-\bar\rho)\sigma-(\edt-\tau-\bar\tau')\kappa=\Psi_0 .\label{E3}
\end{equation}
Applying the commutators $[\tho,\tho']$ and $[\edt,\edt']$  \eqref{eq:  commutators} to $\Phi_0$, and using the Ricci identities~\eqref{eq:  Ricci} and~\eqref{eq:  M}, gives
\begin{multline}
(\tho'-\bar\rho')(\tho\Phi_0+2\kappa\Phi_1)-(\edt'-\bar\tau)(\edt\Phi_0+2\sigma\Phi_1)
\\=3(\rho\,\tho'\Phi_0-\tau\,\edt'\Phi_0
+(\Psi_2-2\Lambda)\Phi_0)+\Phi_2\Psi_0-4\Phi_1\Psi_1 \label{E4}
\end{multline}

These equations \eqref{E1}--\eqref{E4} hold in any tetrad for any Einstein--Maxwell system.  
Their linearisation on an aligned type-D background (i.e. Carter or PD) leads to the following equations:
\begin{subequations}\label{eq:  lin}
\begin{align}
&(\edt'-\tau')\Psi^{(1)}_0-(\tho-4\rho)\Psi^{(1)}_1\notag\\&\hspace{4em}-(3\Psi_2-4|\Phi_1|^2)\kappa^{(1)}=-2\Pb_1\tho\Phi^{(1)}_0 ,\label{lin1}\\
&(\tho'-\rho')\Psi_0^{(1)}-(\edt-4\tau)\Psi_1^{(1)}\notag\\&\hspace{4em}-(3\Psi_2+4|\Phi_1|^2)\sigma^{(1)}=2\Pb_1\edt\Phi^{(1)}_0,\label{lin2}\\
&(\tho-\rho-\bar\rho)\sigma^{(1)}-(\edt-\tau-\bar\tau')\kappa^{(1)}-\Psi^{(1)}_0=0,\label{lin3}\\
&2\Phi_1\big[(\tho'+2\rho'-\bar\rho')\kappa^{(1)}-(\edt'+2\tau'-\bar\tau)\sigma^{(1)}+2\Psi^{(1)}_1\big]\notag\\&\hspace{12em}=-\mathcal M\Phi^{(1)}_0,\label{lin4}
\end{align}
\end{subequations}
where the background quantities are those of~\eqref{eq:  bgscalars} and~\eqref{eq:  spinexplicit}, and
\[
\mathcal M=(\tho'-\bar\rho')\tho-(\edt'-\bar\tau)\edt-3\rho\tho'+3\tau\edt'-3\Psi_2+6\Lambda
\]
is the spin-1 Teukolsky-type operator on an object of weight $\{2,0\}$. By ``priming'' the GHP quantities and operators, one is able to obtain the analogous equations for $\Psi_4^{(1)},~\Psi_3^{(1)},~\kappa'^{(1)},~\sigma'^{(1)},~\Phi_2^{(1)}$.

\paragraph*{Gauge freedom.} 
We have the freedom to perform null rotations on the tetrad, 
\begin{align}
{\bm k}\to {\bm k}+\bar\zeta{\bm m}+ \zeta \bar {\bm m}+|\zeta|^2{\bm l}\,,\quad
{\bm m}\to {\bm m}+\zeta{\bm l}\,,
\end{align}
with $\zeta$ first order of weight $\{2,0\}$. 
Denoting the change in a quantity by this transformation $\delta_\zeta$, we have the following shifts
\begin{equation}\label{eq:  gauge}
\begin{gathered}
\delta_\zeta \Psi^{(1)}_0=0,\quad \delta_\zeta\Psi^{(1)}_1=3\zeta\Psi_2,\quad \delta_\zeta\Phi^{(1)}_0 =2\zeta\Phi_1,\\
\delta_\zeta\kappa=-(\tho-\rho)\zeta,\qquad \delta_{\zeta}\sigma=-(\edt-\tau)\zeta .
\end{gathered}
\end{equation}
The linearised equations however~\eqref{eq:  lin} are invariant. 
For $\fe\neq0$, $\zeta=-\Phi^{(1)}_0 /(2\Phi_1)$  sets $\Phi^{(1)}_0 =0$  the \emph{phantom gauge}~\cite{chandrasekhar1998mathematical}. 
This does not set the Maxwell perturbation to zero but decouples it from the system.
Then the right hand sides of ~\eqref{lin1}--\eqref{lin4} all vanish, they are four equations for the four unknowns $\Psi^{(1)}_0,\Psi^{(1)}_1,\kappa^{(1)},\sigma^{(1)}$, the same system introduced in \cite{Hintz:2026jiw,Yoshida:2026yly} on the PD background. 
For $e=0$,~\eqref{lin1}--\eqref{lin4} decouple and give the Teukolsky equation~\cite{Teukolsky1973}.

Henceforth, we drop the ${}^{(1)}$ on the perturbed quantities as there should be no ambiguity.

\section{Chiral form of the equations}\label{Sec: chiral}

Before deriving the chiral system we make use of the conformal properties of the system. Recalling that a hatted object denotes an object  in the off-shell Carter spacetime $\hat{g}=\Omega^2 g$, we can make use of the GHP calculus to relate objects to the rescaled PD spacetime. For $\eta=\Omega^s\hat\eta$ of weight $\{p,q\}$, we have
\begin{equation}\label{eq:  opstrans}
\begin{aligned}
\tho\eta&=\Omega^{s+2}(\hat\tho+s\,\hat\tho L)\hat\eta,\\
\tho'\eta&=\Omega^{s}\big(\hat\tho'+(s-p-q)\hat\tho'L\big)\hat\eta,\\
\edt\eta&=\Omega^{s+1}\big(\hat\edt+(s-q)\hat\edt L\big)\hat\eta,\\
\edt'\eta&=\Omega^{s+1}\big(\hat\edt'+(s-p)\hat\edt'L\big)\hat\eta ,
\end{aligned}
\end{equation}
where again $L=\log\Omega$.
One can show that by rescaling the perturbed quantities $\{\Psi_0,\Psi_1,\sigma,\kappa \}$ as
\begin{equation}\label{eq:  recon}
\begin{gathered}
\Psi_0=\Omega^5\hat\Psi_0,\qquad \Psi_1=\frac{\Omega^4\hat\Psi_1}{\sqrt2\,\bar z},\\
\kappa=\sqrt2\,\Omega^3\bar z^2\hat\kappa,\qquad \sigma=\frac{\Omega^2\bar z^2}{z}\hat\sigma,
\end{gathered}
\end{equation}
the $L$ terms cancel identically from \eqref{lin1}--\eqref{lin4}. 
However, the system is not quite conformally invariant as is the case for the Teukolsky equation~\cite{Araneda:2018ezs} without matter, or with the freezing approximation of Dudley and Finley~\cite{dudley1977}, also used in \cite{Gray:2025lwy}. 
The coupling of the Ricci/Maxwell scalars breaks this and so there is still explicit $\Omega$ dependence. 

Consider now the perturbed variables to be eigenfunctions of the Killing directions,
i.e., for any $\hat{Z}\in\{\hat \Psi_0,\hat\Psi_1,\hat\kappa,\hat\sigma\}$ the only dependence on the Killing coordinates is  
\be\label{eq: mode dependence}
\hat Z=e^{i(\omega\eta+{m_{\phi}}\phi)}\tilde{Z}(q,p)\,.
\ee
Further, introduce the notation $K=\omega q^2-{m_{\phi}}$, $H=\omega p^2+{m_{\phi}}$, $u=\sqrt Q$, $v=\sqrt P$, then the {derivatives along}  the principal null directions of the Carter metric act on these modes as:
\begin{equation}\label{eq: PN shift}
\hat k\to\pa_q+\dfrac{iK}{Q},\quad \hat l\to-\dfrac{Q}{2\Sigma}\Big(\pa_q-\dfrac{iK}{Q}\Big),\quad \hat m\to\dfrac1{\sqrt2z}\Big(v\pa_p-\dfrac Hv\Big).
\end{equation}
Thus, we can introduce the same set of operators as in \cite{Hintz:2026jiw,Yoshida:2026yly}:
\begin{align}\label{eq:  mode operators}
\mathcal D_+&=\pa_q+\dfrac{iK}{Q}+\dfrac3{\bar z},\qquad \mathcal L_+=v\pa_p+\dfrac Hv+\dfrac{P'}{v}+\dfrac{3iv}{\bar z},\\
\mathcal D'_-&=\pa_q-\dfrac{iK}{Q}+\dfrac{2Q'}{Q}-\dfrac3{\bar z},\quad \mathcal L_-=v\pa_p-\dfrac Hv-\dfrac{P'}{2v}-\dfrac{3iv}{\bar z},
\end{align}
Using the conformal rescalings~\eqref{eq:  recon} we can turn~\eqref{lin1}--\eqref{lin4} into
\begin{subequations}\label{eq:  abcd}
\begin{align}
\mathcal L_+\hat\Psi_0-\mathcal D_+\hat\Psi_1&=\Big(6\Pi-2\Omega e^2\frac{\bar z}{z^2}\Big)\hat\kappa,\\
Q\mathcal D'_-\hat\Psi_0+\mathcal L_-\hat\Psi_1&=-\Big(6\Pi+2\Omega e^2\frac{\bar z}{z^2}\Big)\hat\sigma,\\
\mathcal D_+\hat\sigma-\mathcal L_-\hat\kappa&=\Omega\frac{z}{\bar z^2}\hat\Psi_0,\\
Q\mathcal D'_-\hat\kappa+\mathcal L_+\hat\sigma&=2\Omega\frac{z}{\bar z^2}\hat\Psi_1 .
\end{align}
\end{subequations}
with $\Pi$ given above in \eqref{eq:  bgscalars}.
Notice the $\Omega$ dependence is very minimal.

The negative spin weights are recovered from primed equations with 
\begin{equation}
    \begin{aligned}
        \Psi_4&=&\frac{\Omega Q^2}{2\bar z^4}\hat\Psi_4, \quad \Psi_3&=-\frac{\Omega^2Q}{\sqrt2{\bar z}^3}\hat\Psi_3,\\
        \kappa'&=&\frac{Q^2}{\sqrt2\Sigma\Omega}\hat\kappa', \quad \sigma'&=-\frac Q{\bar z}\hat\sigma'\,,
    \end{aligned}
\end{equation} these become~\eqref{eq:  abcd} with $(\hat\Psi_0,\hat\Psi_1,\hat\kappa,\hat\sigma)\to(\hat\Psi_4,\hat\Psi_3,\hat\kappa',\hat\sigma')$ and $(\omega,{m_{\phi}})\to(-\omega,-{m_{\phi}})$. 
The origin of these relations is in the time-reversal plus inversion symmetry $(\eta,\phi)\to(-\eta,-\phi)$  of~\eqref{eq:  PD} which maps the Carter tetrad to its GHP prime up to a boost and a spin, reverses the mode, and changes $A\to-A$, which enters only through the charge $\fe$.

Finally, we let
\begin{equation}\label{eq: y}
\begin{gathered}
\psi_\pm=\hat\Psi_0\mp\frac iu\hat\Psi_1,\qquad \chi_\pm=\hat\kappa\mp\frac iu\hat\sigma,\\
y_\pm=Q^{5/4}P^{1/4}\,\big(\psi_\pm,\ 2e\,\chi_\pm\big)^T .
\end{gathered}
\end{equation}

Then the system is equivalent to
\begin{equation}\label{eq: chiral}
(\sD-\sA)\,y_+=\sF\,y_-,\qquad (\bar\sD+\sA)\,y_-=\mathrm{adj}\,\sF\,y_+,
\end{equation}
where, $\sD=u\pa_q+iv\pa_p$, for an invertible $\sF$ the adjugate matrix is $\mathrm{adj}\,\sF=(\det \sF)\,\sF^{-1}$ , and
\begin{equation}\label{eq: AF}
\begin{gathered}
\sA=\Omega e\begin{pmatrix}0&-\dfrac{i\bar z}{z^2}\\[6pt] \dfrac{iz}{\bar z^2}&0\end{pmatrix},
\qquad
\sF=\begin{pmatrix}\mathcal C-i\mathcal B&\dfrac{3i}{e}\Pi\\[6pt] -\dfrac{3ie \Omega z}{\bar z^2}&\mathcal C+i\mathcal B\end{pmatrix},\\
\mathcal C=\frac{iK}{u}+\frac{3u}{\bar z}-\frac{3Q'}{4u},\qquad \mathcal B=\frac Hv+\frac{3iv}{\bar z}+\frac{3P'}{4v}.
\end{gathered}
\end{equation}
The matrix $\sA$ is Hermitian.
When $\Omega=1$, i.e. $\alpha=0$, and the metric functions are Kerr-Newman-(A)dS this is the system of~\cite{Yoshida:2026yly}.

The techniques of \cite{Hintz:2026jiw,Yoshida:2026yly} have a chance to apply if the determinant of $F$ is separable, 
and indeed it is,
\begin{equation}\label{eq: detF}
\begin{gathered}
\det\sF=\Vq(q)+\Vp(p),\\
Q\Vq=-\big(K+\tfrac34iQ'\big)^2+Q\big(6i\omega q-\tfrac34Q''\big),\\
P\Vp=\big(H+\tfrac34P'\big)^2-P\big(6\omega p+\tfrac34P''\big).
\end{gathered}
\end{equation}
We proceed to show that the matrix integrating factor exists for this system following the arguments laid out in \cite{Hintz:2026jiw,Yoshida:2026yly}. 
We find Yoshida's~\cite{Yoshida:2026yly} approach more amenable to the problem at hand.

\section{Matrix $R$-separation}\label{Sec: Separation}

The construction begins from the observation that the determinant of $\sF$ separates as in \eqref{eq: detF}. 
Following \cite{Yoshida:2026yly}, one seeks a matrix $\sX$ to transform the system into separable coordinates, such that $\det \sX=1$, and
\begin{equation}
    \sD \sX=\sA \sX+\sX(\vartheta_p(p)+\vartheta_q(q))\sigma_2\label{X_eq}
\end{equation}
with $\vartheta_q$ and $\vartheta_p$ being functions of only $q$ and $p$ that we will determine later.
We use the Pauli matrices as a basis for $2\times2$ matrices.
That is $\sigma_0=\mathds{1}_2$,
\be
\sigma_1=\begin{pmatrix}
        0&1\\
        1&0
        \end{pmatrix},\quad
\sigma_2=\begin{pmatrix}
        0&-i\\
        i&0
        \end{pmatrix},\quad
\sigma_3=\begin{pmatrix}
        1&0\\
        0&-1
        \end{pmatrix}\,.   
\ee
With these somewhat obscure forms, under the transformation
\be
z_+=-\sX^{-1} y_+\,, z_-=\sX^\dagger y_-\,
\ee
the system becomes
\begin{subequations}
\label{eq: z system}
\begin{align}
    [\sD +(\vartheta_p +\vartheta_q)\sigma_2]z_+&=\sC z_-,  \\
    [\bar\sD -(\vartheta_p +\vartheta_q)\sigma_2]z_-&=\text{adj} \sC \,z_+\,,
\end{align}
\end{subequations}
where
\begin{align}
  \sC&=-\sX^{-1}\sF (\sX^\dagger)^{-1}\nonumber\\
   &=C_0(q)\sigma_0+C_2(q)\sigma_2+\im C_1(p)\sigma_1+\im C_3(p)\sigma_3\,.  \label{C_def}
\end{align}
Notice that 
\be
\det \sF =\det \sC=(C_0^2-C_2^2) + (C^2_1+C_3^2)\,.
\ee
Hence we can immediately identify
\be
\Vq(q)=C_0^2-C_2^2+{\mathsf c}\,,\quad \Vp(p)=C^2_1+C_3^2-{\mathsf c}\,
\ee
for some constant ${\mathsf c}$ to be determined.
The exact form of $C_i$ depends on the choice of $\sX$ which has residual freedom we discuss below.
However the particular combinations $C_0^2-C_2^2$ and $C_1^2+C_3^2$ are invariant.

If we now decompose the system in terms of the eigenvectors of $\sigma_2$
\be
 e_\pm=\tfrac{1}{\sqrt{2}}(1,\pm i)^t, \sigma_2e_\pm=\pm e_\pm,\sigma_1e_\pm=\pm i e_\mp,\sigma_3e_\pm=e_\mp,
\ee
so that
\begin{align}
    z_+&= U_1 e_-+V_2 e_+\,, z_-= V_1e_- +U_2 e_+, 
    \\
    U&=(U_1,U_2)^t, V=(V_1,V_2)^t\,.
\end{align}
Then \eqref{eq: z system} becomes:
\begin{align}
    u\pa_q U&=(\sD_p+\sigma_0\vartheta_q)U+ (C_0-C_2) V\\
    u\pa_q V&=-(\sD_p+\sigma_0\vartheta_q)V+ (C_0+C_2) U
\end{align}
where we have introduced the angular operator
\be\label{eq: angular op}
\sD_p=-i\sigma_3 v\pa_p -C_1\sigma_1-C_3\sigma_3+\vartheta_p\sigma_0\,.
\ee
Finally, writing 
\be
U=R_1(q)
\begin{pmatrix}
    S_1(p)\\
    S_2(p)
\end{pmatrix}
,
V=R_2(q)
\begin{pmatrix}
    S_1(p)\\
    S_2(p)
\end{pmatrix}
\ee
one obtains the coupled separated\footnote{Here by coupled we mean the coupling between functions with indices $1$ and $2$, while by ``separated'' we mean that the equations for $q$- and $p$-dependent functions are distinguished.} equations
\begin{align}\label{eq: R sep}
&\sqrt Q\frac{d}{dq}
\begin{pmatrix}
R_1\\R_2
\end{pmatrix}
=\begin{pmatrix}
\varpi+\vartheta_q&C_0-C_2\\ C_0+C_2&-\varpi-\vartheta_q
\end{pmatrix}
\begin{pmatrix}
R_1\\R_2
\end{pmatrix},
\\
\label{eq: S sep}
&\begin{pmatrix}
-i\sqrt P\frac{d}{dp}+\vartheta_p&-C_1+iC_3\\
-C_1-iC_3&i\sqrt P\frac{d}{dp}+\vartheta_p
\end{pmatrix}
\begin{pmatrix}
S_1\\S_2
\end{pmatrix}
=\varpi
\begin{pmatrix}
S_1\\S_2
\end{pmatrix}.
\end{align}
where $\varpi$ is the shared eigenvalue of the system.
The origin of the shared eigenvalue for $S$ and $R$ is the symmetry operator in the next section, and can be seen from the related fact that only the one angular operator \eqref{eq: angular op} enters the system in this basis.

Thus the problem is reduced to obtaining such a matrix $\sX$, which provides the $R$-separation of $z_\pm$.
We seek an intermediate matrix $\sJ$ which satisfies
\be\label{J_def}
\sJ=\sX\sigma_2\sX^{-1}, \sJ^2=\mathds{1}_2, \sD \sJ=[\sA,\sJ]\,,
\ee
Thus $\sX$ diagonalises $\sJ$ and so can be written in terms of its eigenvectors.

The final two equations of \eqref{J_def}, $\sJ^2=\mathds{1}_2, \sD \sJ=[\sA,\sJ]\,$, in fact, simplify to the set of linear equations
\begin{align}
    &\sD \sJ_{12}=-2 \sA_{12}\sJ_{11},~~~\sD \sJ_{21}=2 \sA_{21}\sJ_{11},\nonumber\\
    &\sD \sJ_{11}=\sA_{12}\sJ_{21}-\sJ_{12}\sA_{21}.
\end{align}

These equations are linear in $\sJ_{ij}$ and can be solved for the coefficients since $u$ and $v$ are known. 
As we want to obtain Yoshida's result as the limit, we substitute the solution in the form
\begin{align}
    \sJ_{11}=-\sJ_{22}=\dfrac{2 e}{z\bar{z}}j_{11},~~~\sJ_{12}=\dfrac{\im }{z^2}j_{12},~~\sJ_{21}=-\dfrac{\im}{\bar{z}^2}j_{21}
\end{align}

This substitution simplifies the corresponding equations, and the solution for $j_{11},~j_{12}$ and $j_{21}$ can then be found as a polynomial in $u,~v,~q$ and $p$.
The result can be given by
\begin{align}
    j_{11}=\dfrac{\mathcal{N}}{\nu |\mu|},~~~j_{12}=\dfrac{\bar{\mu}}{\nu |\mu|}\mathcal{G},~~j_{21}=\dfrac{\mu}{\nu|\mu|}\tilde{\mathcal{G}}
\end{align}
where
\begin{align}
    &\mathcal{N}=v\big(|\mu|^2q-e^2(m+\alpha n q^2)\big)-\im u\big(|\mu|^2p-e^2 (n+\alpha m p^2)\big),\nonumber\\
    &\mathcal{G}=e^2(u+v)^2+e^2(\lambda+\alpha^2K_\mu)z^2\bar{z}^2-\dfrac{\mathcal{Z}^2}{|\mu|^2},\nonumber\\
    &\tilde{\mathcal{G}}=e^2(u-v)^2+e^2(\lambda+\alpha^2K_\mu)z^2\bar{z}^2-\dfrac{\bar{\mathcal{Z}}^2}{|\mu|^2}\\
    &\mathcal{Z}=|\mu|^2 z-e^2(\mu+\alpha n q^2+\im \alpha m p^2),~~~K_\mu=k+\dfrac{e^2n^2}{|\mu|^2},\nonumber\\
    &\nu^2=\Big[|\mu|^2-e^2\Big(\epsilon-\dfrac{2\alpha m n e^2}{|\mu|^2}\Big)\Big]+4 e^4 K_\mu(\lambda+\alpha ^2 K_\mu).\nonumber
\end{align}
The eigenvectors of the matrix $\sJ$ are $\xi_{\pm}$ that are given by
\begin{align}
    \xi_{\pm}=\begin{pmatrix}
        \sJ_{12}\\
        \pm 1-\sJ_{11}
    \end{pmatrix}
\end{align}
that satisfy 
\begin{align}
    \sJ \xi_{\pm}=\pm \xi_{\pm}
\end{align}

The fact that eigenvalues are $\pm1$ follows from the definition of $\sJ$ \eqref{J_def} and from the fact that eigenvalues of $\sigma_2$ are $\pm1$.

Now, when we know the matrix $\sJ$, we can restore the matrix $\sX$ from the eigenvectors of $\sJ$. To do this, we at first notice that as $\sigma_2=\sU\sigma_3 \sU^\dagger$ where $\sU=(e_+,e_-)$. From the other side, we can write
\begin{align}\label{eq: J eigensystem}
    \sJ=(f\xi_+,h\xi_-)\sigma_3(f\xi_+,h\xi_-)^{-1}
\end{align}

By comparing this expression with (\ref{J_def}), we obtain
\begin{align}
    \sX=(f\xi_+,h\xi_-)\sU^\dagger\label{X_sol}
\end{align}
where $f$ and $h$ are generic functions of $q$ and $p$. However, we impose an additional condition on $\sX$, namely that $\det \sX=1$. This condition fixes the function $h$ that is given by
\begin{align}
    h=\dfrac{1}{-2\im \sJ_{12}f}
\end{align}

Thus, we are left with the determination of the function $f$. 
This function can be found from the condition that the matrix $\sX$ satisfies the following conditions
\begin{align}
    \sD \sX=\sA\sX+\vartheta \sX\sigma_2
\end{align}
with $\vartheta$ being separable $\vartheta=\vartheta_q(q)+\vartheta_p(p)$. 
However, the corresponding equation is quite complicated as it requires solving a PDE.
Fortunately, there still exists an additional freedom in the function $f$, we can send 
\be
f\to e^{\gamma}f=e^{\gamma_q+i\gamma_p}f,
\ee
where $\gamma_{q/p}$ are real functions of $q/p$ respectively.
This transformation amounts to sending
\begin{align}
    \sX &\to \sX e^{\gamma\sigma_2}\,, \nonumber\\
    \sC &\to e^{-\gamma\sigma_2}\sC e^{-\bar{\gamma}\sigma_2}\nonumber\\
    C_0\pm C_2&\to e^{\mp2\gamma_q}( C_0\pm C_2)\,,\nonumber\\
    C_1\pm iC_3&\to e^{\mp2i\gamma_p}( C_1\pm \im C_3)\,.
\end{align}
The separation of variables is unchanged, but the individual functions are:
\begin{align}
S_{1,2}\to e^{\pm i\gamma_p} S_{1,2}\,, \quad R_{1,2}\to e^{\pm \gamma_q} R_{1,2}\,,
\nonumber\\
\vartheta_q\to \vartheta_q+u\gamma'_q,\quad\vartheta_p\to \vartheta_p-v\gamma'_p.
\end{align}
We call a specific choice of $f$, a ``frame'' and now fix a particular frame by requiring that $C_2$  and $C_3$ do not depend on $\omega$. 
This condition turns out to be different to the expressions provided by Yoshida in the Kerr--Newman-(A)dS limit~\cite{Yoshida:2026yly}, however we provide an explicit translation between the two.
The nice property of this choice is that it preserves the duality of the PD metric in \eqref{eq:  PD} and \eqref{eq:  metric funcs}.
Namely, it is symmetric under 
\be\label{eq: duality}
q\to ip,\: (m, n)\to (in , -im), \: e^2\to -e^2, \: k\to k+e^2\,,
\ee
which also sends $Q\to P$.

By requiring the corresponding conditions and using the definitions \eqref{C_def}, one obtains 
\begin{align}\label{eq: f mag}
    |f|^2=&\dfrac{\nu}{4v S_q  \mathcal{G}\bar{\mathcal{G}}|\mu|^2}\Big(\mathcal{G}\bar{\mathcal{G}}|\mu|^2 (u p^2+v q^2)+(v q^2-u p^2)\times\nonumber\\
    &\big[4 e^2\mathcal{N}\bar{\mathcal{N}}+2 e |\mu|(\mathcal{N}+\bar{\mathcal{N}})\nu z\bar{z}+|\mu|^2 \nu^2 z^2\bar{z}^2\big] \Big),\\
    \label{eq: f arg}
    \dfrac{f}{\bar{f}}=&\dfrac{\im}{2 u S_p\mathcal{G}|\mu| \bar{\mu}\bar{z}^2}\Big(\mathcal{G}\bar{\mathcal{G}}|\mu|^2(up^2+v q^2)+(v q^2-up^2)\times\nonumber\\
    &\big[4 e^2\mathcal{N}\bar{\mathcal{N}}-2 e|\mu|(\mathcal{N}-\bar{\mathcal{N}})\nu z \bar{z}-|\mu|^2\nu^2 z^2\bar{z}^2\big]\Big).
\end{align}

The first equation defines the absolute value of $f$, while the second one its argument. 
In these expressions, $S_q$ and $S_p$ are given by
\begin{align}
    S_q^2=\nu^2 q^4+4e^4 K_\mu Q,~~~S_p^2=\nu^2p^4+4 e^4 K_\mu P.
\end{align}

Moreover, this choice of $f$ satisfies the equation \eqref{X_eq} with such functions $\vartheta_q$ and $\vartheta_p$:
\begin{align}
    \vartheta_q=\dfrac{\nu q^2}{S_q^2}\Big[e |\mu| K_\mu-\dfrac{e m \tilde{\nu}+2 \alpha e^3 K_\mu n}{|\mu|}q+\dfrac{|\mu|\tilde{\nu}}{2 e}q^2\Big]\\
    \vartheta_p=\dfrac{\nu p^2}{S_p^2}\Big[e |\mu| K_\mu+\dfrac{e n\tilde{\nu}-2 \alpha e^3 K_\mu m}{|\mu|}p-\dfrac{|\mu|\tilde{\nu}}{2 e}p^2\Big]
\end{align}
where
\begin{align}
    \tilde{\nu}^2=|\mu|^2-e^2\Big(\epsilon-\dfrac{2\alpha e^2 m n}{|\mu|^2}\Big).
\end{align}

From here we can use \eqref{eq: J eigensystem} and \eqref{X_sol} to obtain expressions for the $C_i$ coefficients in \eqref{C_def}.
One obtains:
\begin{align}
    C_0&=-\dfrac{\im \omega S_q}{\nu u}+\dfrac{\im {m_{\phi}} \nu (q^2+2 e^2 Q I) }{S_q u}+\dfrac{3 \nu }{4 u S_{q}}(q^2 Q'-4 q Q),\\
    C_1&=\dfrac{\omega S_p}{\nu v}+\dfrac{\im{m_{\phi}} \nu(p^2-2 e^2P I)}{S_p v}+\dfrac{3\nu}{4 v S_p }(p^2 P'-4 p P),\\
    C_2&=\dfrac{2 \im {m_{\phi}} e}{S_q |\mu|}(|\mu|^2 q-e^2(m+\alpha n q^2))+\dfrac{3|\mu|}{4 e S_q}\mathcal{Q},\\
    C_3&=-\dfrac{2 {m_{\phi}} e}{|\mu| S_p}(|\mu|^2 p-e^2(n+\alpha m p^2))-\dfrac{3|\mu|}{4 e S_p}\mathcal{P}.
\end{align}
where $I=\dfrac{\tilde{\nu}}{\nu^2}$, and two functions $\mathcal{Q}$ and $\mathcal{P}$ (that should not be confused with the metric functions $Q$ and $P$) are given by
\begin{align}
    \mathcal{Q}&=\dfrac{e^2 I \nu^2 q(q Q'-4 Q)-2 e^4 K_{\mu} Q'-2 \nu^2 q^3}{(m^2+n^2)q-e^2(m+\alpha n q^2)},\\
    \mathcal{P}&=\dfrac{e^2 I \nu^2 p(p P'-4P)+2 e^4 K_{\mu} P'+2 \nu^2 p^3}{(m^2+n^2)p-e^2(n+\alpha m p^2)}.
\end{align}

The corresponding functions are, as required, dependent only on $q$ and $p$, and we have the explicit separation desired.
For the negative spin weights $\Psi_4,\Psi_3,\kappa',\sigma'$ the same holds with $(\omega,{m_{\phi}})\to(-\omega,-{m_{\phi}})$ in $\sF$, $\sC$ and $\mathsf c$, while $\sA$, $\sJ$, $\sX$ and $\vartheta$ are unchanged.
Moreover, under the duality transform \eqref{eq: duality}, and $(\omega,{m_{\phi}})\to(-\omega,-{m_{\phi}})$ we have
\be
S_q\to S_p, \: \vartheta_q\to\vartheta_p, \: C_0 \to iC_1, \: C_2\to C_3, \: \varpi \to -\varpi\,,
\ee
and we have a map from the radial to angular equations:
\begin{align}
(S_1(p),S_2(p))&\to(R_2(ip),iR_1(ip)), \nonumber\\
(R_1(q), R_2(q))&\to(-iS_2(-iq), S_1(-i q))\,.
\end{align}

As with all choices of frame there are positives and negatives. 
The current frame, with $f$ in \eqref{eq: f mag} and \eqref{eq: f arg}, yields simpler forms for the coefficients and $C_i$'s.
However it may not be necessarily the easiest for numerical integration with small charge, due to the $1/e$ factors in for example $\vartheta_{p/q}$ and $C_{2/3}$.
This is a generic feature of the phantom gauge used in \eqref{eq:  gauge} as we divide by $\Phi_1\propto \fe - i \frg$. 
However, the final solution in the variables $y_\pm$ is independent of the choice of $f$ and so this problem persists in any frame.

\subsection{Reduction to Kerr--Newman-(A)dS}

For Kerr--Newman-(A)dS the parameters $\alpha,n,\frg$ vanish, and $k=a^2$, $\epsilon=1-\lambda a^2$, $\mu=m$, and $\Omega=1$.
The matrix $\sJ$ is independent of the frame freedom of $f\to f e^{\gamma_q+i\gamma_p}$.
However, if one wants to work in a frame that reduces to Yoshida's~\cite{Yoshida:2026yly} in the Kerr--Newman-(A)dS limit, one can choose this by requiring that $C_1$ be independent of $(\omega,{m_{\phi}})$ and $(\pa_\omega - a^2\pa_{m_{\phi}})C_3=0$. 
First, we must make the Boyer--Lindquist coordinates identifications: $q=r$, $p=a\cos\theta$ and $\eta=(t-a\phi_{\rm BL})/\Xi$, $\phi=-\phi_{\rm BL}/(a\Xi)$ with $\Xi=1+\lambda a^2$ and hence Yoshida's modes $e^{-i\omega_{\rm BL}t+i m_{\phi_{\rm BL}} \phi_{\rm BL} }  $ have $\omega=-\Xi\,\omega_{\rm BL}$ and ${m_{\phi}}=a\Xi(a\omega_{\rm BL}-m_{\phi_{\rm BL}})$.
Then Yoshida's gauge is achieved by the following transformation:
\be
e^{2\gamma_q}=\frac{\mathsf D+\mathsf N}{\mathcal S_qS_r},\qquad e^{2i\gamma_p}=-\frac{\mathsf Y+i\mathsf X}{4mp\,\mathcal S_p\sqrt{H_\theta}},
\ee
where 
\be
\begin{aligned}
\mathsf D&=\mathcal S_q^2+a^2\nu^2(q^2+2\fe^2IQ),\quad \mathsf N=2a^2\fe\nu(mq-\fe^2)\,u,\\
\mathsf X&=2\fe\nu m\,p\,(p^2P'-4pP),\\
\mathsf Y&=v\big[\nu^22\fe^2I(p^2P'-4pP)+4\nu^2p^3+4a^2\fe^4P'\big].
\end{aligned}
\ee
Here, $S_r^2=\nu^2(q^2+a^2)^2+4a^2m^2\fe^2Q$ and $H_\theta=m^2P+\Xi^2\fe^2p^2$ are as in~\cite{Yoshida:2026yly}.
This takes~\eqref{eq: R sep} and \eqref{eq: S sep} to Yoshida's system, together with the shift:
\be
\begin{gathered}
\varpi \to \varpi+c_{\rm Y},\qquad c_{\rm Y}=\frac{\nu+\fe^2\Xi}{2m\fe},\\
\alpha_r=\vartheta_q+u\gamma_q'-c_{\rm Y},\qquad \alpha_\theta=\vartheta_p-v\gamma_p'+c_{\rm Y}.
\end{gathered}
\ee
and the change of parameters $\fe\to Q,~m\to M,~\lambda\to \mathscr{l}^{-2}$ (for dS case).
Finally, we show in the next section that its origin corresponds to a symmetry operator.

\section{Symmetry operator for the Chiral system}\label{Sec: Sym Operator}

The existence of the $R$-separated form is mysterious. 
Certainly, the construction of $\sX$ and $\sJ$ is not intuitive beyond the idea of removing the action of the matrix $\sA$ from the chiral system \eqref{eq: chiral}, and trying to find a way to put $\sF$ in a form where it is in parts dependent only on $p$ and only on $q$. 

In order to try to understand this in a bit more detail we take inspiration from similar systems for the Dirac equation~\cite{Kalnins:1986wa} with matrix potential~\cite{Carignano_2011}.
Collect the variables $\sz=(z_+,z_-)$ and write~\eqref{eq: z system} as $\mathbb Dz=0$, with, in $2\times2$ blocks,
\begin{equation}\label{eq:bbD}
\begin{gathered}
\mathbb D=u\pa_q+i\Gamma\,v\pa_p-\mathbb V,\\ 
\Gamma=\begin{pmatrix}
\mathds{1}_2&0\\0&-\mathds{1}_2
\end{pmatrix},\qquad 
\mathbb V=
\begin{pmatrix}
-(\vartheta_p+\vartheta_q)\sigma_2&\sC\\ 
\mathrm{adj}\,\sC&(\vartheta_p+\vartheta_q)\sigma_2
\end{pmatrix}.
\end{gathered}
\end{equation}
On shell of the chiral equations we are free to replace the operator as $i\Gamma v\pa_p=\mathbb D-u\pa_q+\mathbb V$. 
Hence any first-order operator, up to a multiple of $\mathbb D$,  can be written in the form 
\be
\sK=\mathsf S\,u\pa_q+\mathsf W\,,
\ee
where we denote the arbitrary $4\times4$ matrices as $\mathsf S(q,p)$, $\mathsf W(q,p)$. 
We find that in the separated frame
\be\label{eq: K op}
\mathsf S=\text{diag}(-\sigma_2, +\sigma_2),
\mathsf W=
\begin{pmatrix}
    -\vartheta_q\sigma_0& C_2\sigma_0 +C_0\sigma_2\\
    C_2\sigma_0 -C_0\sigma_2&-\vartheta_q\sigma_0\\
\end{pmatrix}
\ee
and that
\be
\mathbb D( \mathbb K (\sz))=\tilde{\mathbb K} (\mathbb D (\sz))\,,
\ee
where $\tilde{\mathbb K}$ is \eqref{eq: K op} with each sign of  $C_i$ reversed. 
For solutions of \eqref{eq: R sep} and \eqref{eq: S sep}  one obtains
\be
\mathbb K z=\varpi z\,,
\ee
thus the separation constant is the eigenvalue of $\mathbb K$, cf. the angular operator introduced above in \eqref{eq: angular op}. 
To change the eigenvalue by a constant e.g. to recover Yoshida's system one can always add a multiple of the identity to $\mathbb K$.
Rotating $\mathbb K$ by $G=\text{diag}(-\sX,(\sX^\dagger)^{-1})$ then one obtains a symmetry operator for the $y_\pm$ variables $G\mathbb K G^{-1}$ (again up to multiples of the equations of motion).

This completes our discussion of the separation of the gravitational radiative system for PD spacetimes.

\section{Summary and Outlook} \label{Sec: Outlook}

In this work we have applied and generalised the method of Hintz~\cite{Hintz:2026jiw}, Yoshida~\cite{Yoshida:2026yly}, and Zhao~\cite{Zhao:2026gen}, to decouple and separate the radiative system of gravitational perturbation on the full class of Pleba\'nski--Demia\'nski spacetimes. 
The origin of the separability was also shown to stem from a first order symmetry operator reminiscent of those for the Dirac equation~\cite{McLenaghan1979,Carignano_2011}.
The underlying geometric structure of this operator is an open question, however we suspect it must be inherited from the symmetries of the underlying spacetime, namely, the conformal Killing--Yano tensor or equivalently the conformal Killing spinor.

As this is the most general type D spacetime with aligned electromagnetic fields there is just one possible extension of the method. 
Namely, to the recently discovered Ovcharenko--Podolsk\'y spacetimes~\cite{Podolsky:2025tle,Ovcharenko:2025cpm} which have non-aligned electromagnetic matter fields.
These are unique up to a few minor assumptions \cite{Ovcharenko:2026pow,Nakajima:2026lkr}, and so, would close the story for this particular methodology.
However, the non-alignment of the matter fields means that the phantom gauge is no longer possible. 
A new gauge is needed, nevertheless we believe this is likely possible and is work in progress. 

On the other hand, the separation we have here will prove useful for the analysis of quasi-normal modes and perhaps mode stability of Pleba\'nski--Demia\'nski spacetimes. 
The difficultly here lies in the identification of the correct asymptotic and near horizon behaviour.
Moreover, the conical singularity introduced by the both NUT and acceleration parameters will affect the discussion of this type of problem.
In addition, it is unclear how the freedom to scale the separated functions $R_{1,2},S_{1,2}$, via a choice of frame, affects the asymptotic behaviour of the solutions which may pose numerically difficulties (although the full variables $y_\pm$ are frame independent).
As such we leave this to future work.

Finally, and more speculatively, one of the fundamental obstruction to separation of gravitational perturbations in higher dimensions is that both the Maxwell and curvature scalar perturbations do not decouple even in the vacuum case~\cite{Durkee:2010qu}.
Perhaps some higher dimensional generalisation of this chiral form, and symmetry operator, exists and thereby overcomes this obstruction.

\acknowledgments
The work of FG is supported by the Schwinger Foundation. H.O. is supported by the Czech Science Foundation Grant No.~GA\v{C}R 26-22381S, and by the Charles University Grant No.~GAUK 260325. 
{We thank J. Podolsk\'y and D. Kubiz\v{n}\'ak for valuable comments on the manuscript.}
During this work, {\sf  ANTHROP\textbackslash C's Claude} was used to conduct some technical calculations. 
Conceptualization of the work and, realizing the ideas on separation, were done by the Authors. 
All results generated by Claude were verified by, and are the sole responsibility, of the Authors. 
The calculation of various GHP quantities was facilitated with {\sf xAct}  \cite{xAct} for {\sf Wolfram Mathematica} \cite{Mathematica} and, in particular, the spinframes package \cite{Aksteiner:2022fmf}.

\appendix

\section{Bianchi, Ricci, and commutator identities }\label{App: Exact eqs}

We collect here the Bianchi, Ricci and commutator identities for the GHP formalism in a general background used to derive the radiative system in the conventions of \cite{Stephani:2003tm}.

First, the Ricci identities:
\begin{subequations}\label{eq:  Ricci}
\begin{align}
\tho\rho-\edt'\kappa&=\rho^2+\sigma\bar\sigma-\bar\kappa\tau-\tau'\kappa+\Phi_{00},\\
\tho\sigma-\edt\kappa&=(\rho+\bar\rho)\sigma-(\tau+\bar\tau')\kappa+\Psi_0,\\
\tho\tau-\tho'\kappa&=(\tau-\bar\tau')\rho+(\bar\tau-\tau')\sigma+\Psi_1+\Phi_{01},\\
\edt\rho-\edt'\sigma&=(\rho-\bar\rho)\tau+(\bar\rho'-\rho')\kappa-\Psi_1+\Phi_{01},\\
\tho'\rho-\edt'\tau&=\rho\bar\rho'+\sigma\sigma'-\tau\bar\tau-\kappa\kappa'-\Psi_2-2\Lambda,
\end{align}
\end{subequations}
with their primes. 

Then, the Bianchi identities:
\begin{subequations}\label{eq:  Bianchi}
\begin{align}
&(\edt'-\tau')\Psi_0-(\tho-4\rho)\Psi_1-3\kappa\Psi_2=(\edt-\bar\tau')\Phi_{00}\nonumber\\
&\quad-(\tho-2\bar\rho)\Phi_{01}+2\sigma\Phi_{10}-2\kappa\Phi_{11}-\bar\kappa\Phi_{02},\\
&(\tho'-\rho')\Psi_0-(\edt-4\tau)\Psi_1-3\sigma\Psi_2=(\edt-2\bar\tau')\Phi_{01}\nonumber\\
&\quad-(\tho-\bar\rho)\Phi_{02}+2\sigma\Phi_{11}-2\kappa\Phi_{12}+\bar\sigma'\Phi_{00}.
\end{align}
\end{subequations}
Finally, the commutators on weight $\{p,q\}$ objects are
\begin{subequations}\label{eq:  commutators}
    \begin{align}
[\tho,\tho']&=(\bar\tau-\tau')\edt+(\tau-\bar\tau')\edt'\nonumber\\
&\quad-p(\kappa\kappa'-\tau\tau'+\Psi_2+\Phi_{11}-\Lambda)\nonumber\\
&\quad-q(\bar\kappa\bar\kappa'-\bar\tau\bar\tau'+\bar\Psi_2+\Phi_{11}-\Lambda),
\\
[\edt,\edt']&=(\bar\rho'-\rho')\tho+(\rho-\bar\rho)\tho'\nonumber\\
&\quad+p(\rho\rho'-\sigma\sigma'+\Psi_2-\Phi_{11}-\Lambda)\nonumber\\
&\quad-q(\bar\rho\bar\rho'-\bar\sigma\bar\sigma'+\bar\Psi_2-\Phi_{11}-\Lambda),
\\
[\tho,\edt]&=\bar\rho\edt+\sigma\edt'-\bar\tau'\tho-\kappa\tho'\nonumber\\
&\quad-p(\rho'\kappa-\tau'\sigma+\Psi_1)-q(\bar\sigma'\bar\kappa-\bar\rho\bar\tau'+\Phi_{01}),
\end{align}
\end{subequations}
with their primes and conjugates.

\section{Killing spinor and type D}\label{App: K S}

In this appendix we review the notion and implication of the existence of a conformal Killing spinor (CKS). 
We follow the conventions of ~\cite{Stephani:2003tm} (up to a reordering of the coordinates) which introduces some minus signs relative to the Penrose--Rindler conventions~\cite{PenroseRindler1,PenroseRindler2}.
A spin dyad $(o^A,\iota^A)$ with \mbox{$o_A\iota^A=1$} is used to represent the null tetrad $k=o\bar o$, $l=\iota\bar\iota$, $m=o\bar\iota$, with $g(k,l)=-1$, $g(m,\bar{m})=1$.
Formally a soldering form is used to convert spinor to spacetime indices but we do not need that here. 
The numbers of the curvature and Maxwell scalars in \eqref{eq: Weyl coeffs}, \eqref{eq: Ricci coeffs}, and \eqref{eq: max scalars} correspond to the number of $\iota^A$ contractions in the corresponding spinorial object, e.g. $\Phi_0=\Phi_{AB}o^Ao^B, \Phi_{01}=\Phi_{AA'BB'}o^Ao^B\bar o^{A'}\bar\iota^{B'}$, etc.

A CKS is a symmetric spinor $k_{AB}$ which satisfies the following equation:
\begin{equation}\label{eq: KSeq}
\nabla_{A'(A}k_{BC)}=0 .
\end{equation}
The equation is conformally invariant, with $k_{AB}$ of conformal weight one~\cite{PenroseRindler2}.
From $k_{AB}$ one constructs the following 2-form 
\be
f_{AA'BB'}=k_{AB}\epsilon_{A'B'}+\bar k_{A'B'}\epsilon_{AB}
\ee
This and its dual are both conformal Killing--Yano forms~\cite{PenroseRindler2}. 
Typically, $h=\star f$ matches the CKY tensor in \eqref{eq:  CKY}. 
If $k_{AB}k^{AB}\neq0$ we say the CKS is non-degenerate, in which case a frame rotation can be used to set $k_0=0=k_2$ and so
\be
k_{AB}=-2\chi\,o_{(A}\iota_{B)}\,.
\ee 

The defining equation \eqref{eq: KSeq} is an overdetermined PDE. 
By differentiating and using the Ricci identities in the previous section one obtains the following condition on the Weyl spinor~\cite{PenroseRindler2}:
\be
\Psi_{(ABC}{}^{E}k_{D)E}=0 .
\ee
Contracting this with arbitrary spinors $\xi^A=x o^A+y \iota ^A$ leads to the following polynomial relation
\be
\chi\sum_n\tbinom4n(2-n)\Psi_n\,x^{4-n}y^n=0 ,
\ee
which implies $\Psi_0=\Psi_1=\Psi_3=\Psi_4=0$.
Hence the spacetime is of type~D, and the principal null directions are aligned with $k_{AB}$.
Finally, projecting the components of \eqref{eq: KSeq} on the GHP basis leads to a Goldberg--Sachs-like shear-free null congruence
\begin{equation}\label{eq:KSghp}
\begin{gathered}
\chi\kappa=\chi\sigma=\chi\kappa'=\chi\sigma'=0,\\
\tho\chi=-\rho\chi,\quad \edt\chi=-\tau\chi,\quad \tho'\chi=-\rho'\chi,\quad \edt'\chi=-\tau'\chi ,.
\end{gathered}
\end{equation}
Moreover, the second line of the equations determines $\chi$ up to a constant. 
For vacuum type D, $\chi\propto\Psi_2^{-1/3}$~\cite{Walker:1970un}.

\bibliography{ref.bib}
\bibliographystyle{apsrev4-1}
\end{document}